\documentclass[a4paper,11pt]{article}

\usepackage[left=2.5cm,right=2.5cm,top=2.5cm,bottom=2.5cm]{geometry}
\usepackage{graphicx,amssymb,amsmath,amsthm,mathrsfs,setspace,subcaption,cite,authblk,float,cancel}

\usepackage[colorlinks=true,bookmarks=true]{hyperref}
\usepackage{orcidlink} 

\usepackage{mathtools}

\DeclarePairedDelimiterX\braket[2]{\langle}{\rangle}{#1 \delimsize\vert #2}

\hypersetup{citecolor=blue}
\newcommand{\dif}{\mathrm{d}}
\newcommand{\Eqref}[1]{(\ref{#1})}
\newcommand{\half}{\frac{1}{2}}
\newcommand{\expo}[1]{\mathrm{e}^{#1}}
\newcommand{\brac}[1]{\left(#1 \right)}
\newcommand{\sbrac}[1]{\left[#1\right]}

\newif\ifanswers

\newcommand{\RefOne}[1]{
  \ifanswers
   {\color{blue}{#1}}
  \else
   {#1}
  \fi
}

\numberwithin{equation}{section}
 
\begin{document}

\title{Weyl-type solutions with multipolar scalar fields}

\author[1]{Yen-Kheng Lim \orcidlink{0000-0002-0907-6904} \footnote{Email: yenkheng.lim@gmail.com, yenkheng.lim@xmu.edu.my}} %

\affil[1]{\normalsize{\textit{School of Mathematics and Physics, Xiamen University Malaysia, Jalan Sunsuria, 43900 Sepang, Malaysia}}}

\date{\normalsize{\today}}

\renewcommand\Authands{, and }

\maketitle

 \begin{abstract}
  A class of solutions in $d$-dimensional Einstein gravity minimally coupled to a massless scalar field is studied, where the spacetime metric is of a generalized Weyl form with $d-2$ commuting Killing vectors. In addition to the procedure to generate scalar multipolar fields, a $SO(2)$ symmetry can be exploited to generate further solutions. A particular result of this procedure is a solution that contains the scalar counterpart of the Schwarzschild--Melvin and the Fisher--Janis--Newman--Winicour solutions as particular limits. Furthermore, a Harrison-type transformation can also be performed to generate solutions with magnetic fields. Using this transformation we obtain a solution with magnetic and scalar fields present and contains both magnetic and scalar counterparts of Schwarzschild--Melvin as limits.
 \end{abstract}

\section{Introduction} \label{sec_intro}

In Ref.~\cite{Cardoso:2024yrb}, Cardoso and Nat{\'a}rio presented an exact solution to the Einstein--scalar equation which was described as the \emph{scalar counterpart to the Melvin solution}. Recall that the Melvin solution \cite{Bonnor:1954,misra1962some,Melvin:1963qx} describes an axisymmetric bundle of magnetic field lines held together under its own gravity. The scalar counterpart is similarly an axisymmetric solution, where the field present is a scalar rather than a magnetic field. Cardoso and Nat{\'a}rio also showed that a black hole can be included in the solution, and it is convenient to refer to this as the \emph{scalar counterpart to the Schwarzschild--Melvin solution.} The magnetic version of the Schwarzschild--Melvin solution was given by Ernst in \cite{Ernst:1976mzr}.

Subsequently, it was pointed out by Herdeiro \cite{Herdeiro:2024oxn} that the scalar counterpart to the Schwarzschild--Melvin solution is a particular case of spacetimes immersed in a scalar multipolar universe. The procedure used to construct such solutions can be traced back to Eri\c{s} and G\"{u}rses \cite{Eris:1976xj}. When the metric is written in the canonical Weyl form \cite{Weyl:1917}, a scalar field obtained from a solution to the Laplace equation can be superposed to an existing vacuum solution, thus dressing the spacetime with a massless scalar hair. As a general solution to the Laplace equation can be written in a multipole expansion, solutions of different multipoles can be considered. The scalar counterparts to Schwarzschild--Melvin are, in particular, the dipole case with a growing radial part. Since then, this procedure has been applied to find other solutions with massless scalar fields, among them rotating solutions in four and five dimensions \cite{Barrientos:2025abs}, as well as accelerating, rotating, and charged solutions \cite{Stelea:2025ppj}.

In this paper, we consider the procedure of adding scalar fields in the context of generalized Weyl solutions according to a construction by Emparan and Reall \cite{Emparan:2001wk}. Suppose that a $d$-dimensional spacetime has $d-2$ commuting Killing vectors. This can be brought to a form that generalises Weyl's four dimensional canonical form. We will see that the procedure of \cite{Eris:1976xj} to add a scalar field to a vacuum solution continues to hold, and therefore vacuum solutions in the generalized Weyl form can be extended to include scalar multipolar hair. In this way, the four and five-dimensional scalar counterparts to the Schwarzschild--Melvin spacetime is obtained. In five dimensions this procedure can also be applied to the static black ring \cite{Emparan:2001wk}, thus giving a black ring with scalar multipolar hair.

A small modification of the generalized Weyl solution further reveals a $SO(2)$ symmetry in which we can apply Buchdahl's transformation \cite{Buchdahl:1956zz,Buchdahl:1959nk}. When applied to the Schwarzschild solution, it was one the ways to easily derive the Fisher--Janis--Newman--Winicour (FJNW) \cite{Fisher:1948yn,Janis:1968zz} solution, which describes a spherically-symmetric static spacetime with a naked singularity. A detailed study of the properties of the FJNW solution was done in \cite{Abdolrahimi:2009dc}. Recently, the accelerating extension to the FJNW solution was given in \cite{Anjomshoa:2025tdl}. Here we will apply this $SO(2)$ transformation to the black holes with scalar multipolar solutions to obtain a generalisation which contain the previous solutions as particular cases. The decaying multipole solutions preserves some features of the FJNW solution, namely that it has a naked curvature singularity in place of a horizon, and it is asymptotically flat. So the FJNW solution extended with decaying multipoles may be of physical interest as additional candidates in observational searches of naked singularities \cite{Virbhadra:2002ju,Virbhadra:2007kw,Gyulchev:2008ff,Sahu:2012er,Chen:2023uuy}.

With the same modified Weyl solution, one can also include electromagnetic fields to the problem under an appropriate ansatz. In this case a Harrison-type transformation \cite{Harrison:1968} can be applied to magnetize a solution. With this we obtain a single solution which contain \emph{both} magnetic and scalar counterparts of the Schwarzschild--Melvin spacetime.

The rest of the paper is organized as follows. In Sec.~\ref{sec_SolnGen} we review the generalized Weyl form of the metric and derive the equations of motion under Einstein-scalar gravity. With the equations of motion, it is shown how the methods of \cite{Eris:1976xj} and \cite{Buchdahl:1956zz,Buchdahl:1959nk} can be applied to generate new solutions with massless scalar fields. These procedures are then applied in Sec.~\ref{sec_Applications} to various spacetimes that can be cast into the generalized Weyl form, notably the Schwarzschild--Tangherlini solutions, C-metric, as well as the static black ring. In Sec.~\ref{sec_Analysis} we pick out solutions of potential physical interest and analyze some of its properties. In Sec.~\ref{sec_EM} we include electromagnetic fields into the problem and show that the Harrison-type transformation can be easily applied here. A solution containing both magnetic and scalar versions of the Schwarzschild--Melvin spacetime is obtained and some basic properties are studied. Conclusions and closing remarks are given in Sec.~\ref{sec_Conclusion}. Throughout this paper we work in geometric units where $c=G_N=1$, where $c$ is the speed of light and $G_N$ is the Newton's constant in $d$ dimensions. Our convention for Lorentzian signature is $(-,+,\ldots,+)$.

\section{Generating solutions with a scalar field} \label{sec_SolnGen}

Our framework is the Einstein--scalar gravity in $d$ dimensions with the action
\begin{align}
 I=\frac{1}{16\pi}\int\dif^dx\sqrt{-g}\brac{R-\brac{\nabla\varphi}^2},
\end{align}
where $\varphi$ is a massless scalar field, with the notation $\brac{\nabla\varphi}^2=g^{\mu\nu}\partial_\mu\varphi\partial_\nu\varphi$ and $g$ denoting the metric determinant. Extremising the action with respect to the metric and scalar field leads to the equations of motion
\begin{align}
 R_{\mu\nu}=\partial_\mu\varphi\partial_\nu\varphi,\quad\nabla^2\varphi=0, \label{EinsteinScalar_EOM}
\end{align}
where $\nabla^2=g^{\mu\nu}\nabla_\mu\nabla_\nu=\frac{1}{\sqrt{-g}}\partial_\mu\brac{\sqrt{-g}g^{\mu\nu}\partial_\nu}$ is the Laplace operator on the spacetime.

For the metric, we consider the generalised Weyl form of Emparan and Reall \cite{Emparan:2001wk}, in which we take spacetimes that carry $d-2$ orthogonal commuting Killing vectors. Denote the Killing vectors by $\xi_{(1)},\ldots,\xi_{(d-2)}$. Then the theorem in \cite{Emparan:2001wk} states that for each $i$-th Killing vector, if (i) $\xi_{(1)}^{[\mu_1}\cdots\xi_{(d-2)}^{\mu_{d-2}}\nabla^\nu\xi^{\rho]}_{(i)}=0$ at at least one point of spacetime, and (ii) $\xi^\nu_{(i)}{R_\nu}^{[\rho}\xi^{\mu_1}_{(1)}\cdots\xi_{(d-2)}^{\mu_{d-2}]}=0$, then the two-surfaces orthogonal to the $\xi_{(i)}$'s are integrable. Condition (i) is satisfied if at least one of the $\xi_{(i)}$'s is an axial Killing vector. For condition (ii), it may be satisfied if $R_{\mu\nu}=0$ (vacuum spacetime) or $\xi_{(i)}^\mu {R_\mu}^\nu=f\xi_{(i)}^\nu$ for some scalar function $f$. For the Einstein--scalar system, this condition is met if $\xi^\mu_{(i)}\partial_\mu\varphi=0$. (So $f=0$ in the second case.) That is, if $\varphi$ is constant along the Killing directions.

When the conditions are satisfied, coordinates $x^\mu=(y^1,\ldots,y^{d-2},\rho,z)$ can be chosen such that each $y^i$ are aligned along the Killing directions ($\xi_{(i)}=\partial_{y^i}$) and all metric functions are independent of them. The metric then takes the form
\begin{align}
 \dif s^2&=\sum_{j=1}^{d-2}\epsilon_j\expo{2U_j}\brac{\dif y^j}^2+\expo{2\nu}\brac{\dif\rho^2+\dif z^2}, \label{Weyl_metric}
\end{align}
where $\epsilon_j=-1$ if the $j$-th Killing vector is time-like, and $\epsilon_j=+1$ if it is space-like. In this paper we consider spacetimes of Lorentzian signature so all but one of the $\epsilon_j$'s are positive. The functions $U_1,\ldots,U_{d-2}$, and $\nu$ depend only on $\rho$ and $z$. Here we use lowercase Latin indices $i,j,k,\ldots$ to label the coordinates $y^i$ along the Killing directions. The coordinate $\rho$ is chosen such that $\rho^2$  equals the minus determinant of the $(ij)$-part of the metric, hence we have the constraint
\begin{align}
 \sum_{j=1}^{d-2}U_j=\ln\rho, \label{Weyl_constraint}
\end{align}
up to an arbitrary additive constant. The functions $U_j(\rho,z)$ as well as the scalar field $\varphi(\rho,z)$ depend only on the coordinates $\rho$ and $z$. The scalar equation, along with the components $R_{ij}=\partial_i\varphi\partial_j\varphi=0$, $R_{\rho\rho}-R_{zz}=(\partial_\rho\varphi)^2-(\partial_z\varphi)^2$, and $R_{\rho z}=\partial_\rho\varphi\partial_z\varphi$ of the Einstein equation give the equations of motion
\begin{subequations} \label{Weyl_EOM}
\begin{align}
 \vec{\nabla}^2U_j&=0,\quad\vec{\nabla}^2\varphi=0,\label{Weyl_Laplacian}\\
 \partial_\rho\nu&=-\frac{1}{2\rho}+\frac{\rho}{2}\sbrac{\sum_{j=1}^{d-2}\brac{\partial_\rho U_j}^2-\brac{\partial_zU_j}^2}+\frac{\rho}{2}\sbrac{(\partial_\rho\varphi)^2-(\partial_z\varphi)^2},\label{Weyl_nu_rho}\\
 \partial_z\nu&=\rho\sum_{j=1}^{d-2}\partial_\rho U_j\partial_zU_j+\rho\partial_\rho\varphi\partial_z\varphi,\label{Weyl_nu_z}
\end{align}
\end{subequations}
where $\vec{\nabla}^2=\frac{1}{\sqrt{\sigma}}\partial_a\brac{\sqrt{\sigma}\sigma^{ab}\partial_b}=\partial_z^2+\partial_\rho^2+\frac{1}{\rho}\partial_\rho$ is the axisymmetric Laplace operator on an auxiliary flat Euclidean 3-space with (unphysical) metric
\begin{align}
 \dif s^2_{\mathbb{R}^3}=\sigma_{ab}\dif X^a\dif X^b=\dif\rho^2+\rho^2\dif\Phi^2+\dif z^2.\label{R3_aux}
\end{align}
This system of equations determines the metric functions and the scalar field. However there still remains the component $R_{\rho\rho}+R_{zz}=(\partial_\rho\varphi)^2+(\partial_z\varphi)^2$ of the Einstein equation, which is
\begin{align}
 -2\brac{\partial_\rho^2\nu+\partial_z^2\nu}&=-\frac{1}{\rho^2}+\sum_{j=1}^{d-2}\sbrac{(\partial_\rho U_j)^2-(\partial_z U_j)^2} + (\partial_\rho\varphi)^2-(\partial_z\varphi)^2. \label{Weyl_indep}
\end{align}
This can be verified to be consistent with Eqs.~\Eqref{Weyl_Laplacian}--\Eqref{Weyl_nu_z}. To see this, we differentiate Eqs.~\Eqref{Weyl_nu_rho} and \Eqref{Weyl_nu_z} with respect to $\rho$ and $z$ respectively, and adding them up to find
\begin{align*}
 \partial_\rho^2\nu+\partial_z^2\nu&=\frac{1}{2\rho^2}+\sum_{j=1}^{d-2}\sbrac{\half(\partial_\rho U_j)^2-\half(\partial_z U_j)^2+\half(\partial_\rho\varphi)^2-\half(\partial_z\varphi)^2}\\
  &\hspace{2cm}+\rho\sbrac{\sum_{j=1}^{d-2}\partial_\rho U_j(\partial_\rho^2 U_j+\partial_z^2U_j)+\partial_\rho\varphi(\partial_\rho^2\varphi+\partial_z^2\varphi)}.
\end{align*}
From Eq.~\Eqref{Weyl_Laplacian}, we have $\partial_\rho^2U_j+\partial_z^2U_j=-\frac{1}{\rho}\partial_\rho U_j$ and $\partial_\rho^2\varphi+\partial_z^2\varphi=-\frac{1}{\rho}\partial_\rho\varphi$ which can be used to eliminate the second derivatives of $U_j$ and $\varphi$ above and Eq.~\Eqref{Weyl_indep} is recovered. In other words, solutions to Eqs.~\Eqref{Weyl_Laplacian}--\Eqref{Weyl_nu_z} will also solve \Eqref{Weyl_indep}

To reveal the symmetry which will allow Buchdahl transformations, let us pick out one of the $U_j$'s (we choose the last one, $U_{d-2}$) and perform a change of variables
\begin{align}
 U_{d-2}&=U,\quad U_k=V_k-\frac{U}{d-3},\quad k=1,\ldots,d-3,\nonumber\\
 \nu&=\gamma-\frac{U}{d-3}.
\end{align}
we then express the solution in terms of a new set of functions $U$, $V_1,\ldots, V_{d-3}$, and $\gamma$. In terms of the new variables the metric now takes the form
\begin{align}
 \dif s^2&=\epsilon\expo{2U}\dif\sigma^2+\expo{-\frac{2U}{d-3}}\sbrac{\sum_{k=1}^{d-3}\epsilon_k\expo{2V_k}\brac{\dif y^k}^2+\expo{2\gamma}\brac{\dif\rho^2+\dif z^2}}, \label{DimRed_metric}
\end{align}
where $\epsilon=\epsilon_{d-2}$ and $\sigma=y^{d-2}$. The determinant constraint \Eqref{Weyl_constraint} is now
\begin{align}
 \sum_{k=1}^{d-3}V_k=\ln\rho, \label{DimRed_constraint}
\end{align}
up to an additive constant. In particular, $U=U_{d-2}$ is no longer involved in the constraint. In the case $\epsilon=-1$ and $d=4$, we simply have $V_1=\log\rho$ and
recover the usual Weyl form
\begin{align*}
 \dif s^2=-\expo{2U}\dif t^2+\expo{-2U}\sbrac{\rho^2\dif\phi^2+\expo{2\gamma}\brac{\dif\rho^2+\dif z^2}}
\end{align*}
with $\sigma=t$ and $y^1=\phi$.

Further rescaling the scalar field by
\begin{align}
 \varphi=\sqrt{\frac{d-2}{d-3}}\psi, \label{DimRed_scalar}
\end{align}
the equations of motion \Eqref{Weyl_EOM} in terms of the new variables now take the form
\begin{subequations}\label{DimRed_EOM}
\begin{align}
 \vec{\nabla}^2U&=0,\quad\vec{\nabla}^2V_k=0,\quad\vec{\nabla}^2\psi=0,\\
 \partial_\rho\gamma&=-\frac{1}{2\rho}+\frac{\rho}{2}\sum_{k=1}^{d-3}\sbrac{\brac{\partial_\rho V_k}^2-\brac{\partial_z V_k}^2}+\frac{d-2}{d-3}\frac{\rho}{2}\sbrac{\brac{\partial_\rho U}^2-\brac{\partial_zU}^2+\brac{\partial_\rho\psi}^2-\brac{\partial_z\psi}^2},\\
 \partial_z\gamma&=\rho\sum_{k=1}^{d-3}\partial_\rho V_k\partial_z V+\frac{d-2}{d-3}\rho\brac{\partial_\rho U\partial_zU+\partial_\rho\psi\partial_z\psi},
\end{align}
\end{subequations}
where the $V_k$'s are subject to the constraint \Eqref{DimRed_constraint}.

As a brief aside, note the following observation: In the vacuum case ($\psi=0$), Eq.~\Eqref{DimRed_EOM} becomes equivalent to \Eqref{Weyl_EOM}, but in one dimension lower, with $U=\sqrt{\frac{d-3}{d-2}}\varphi$ playing the role of the scalar field in the lower dimensional equations. This is essentially because the metric \Eqref{DimRed_metric} is written in a form in which one can perform a dimensional reduction of the $\sigma$ direction to yield a $(d-1)$-dimensional generalised Weyl metric, which is the expression in the square brackets of \Eqref{DimRed_metric}. In this way, a vacuum gravity in $d$ dimensions has been dimensionally reduced to a $(d-1)$-dimensional gravity with a massless scalar field. This procedure can be repeated so that a $d$-dimensional vacuum gravity can be reduced to a $(d-m)$-dimensional spacetime with $m$ distinct scalar fields.

The equations of motion \Eqref{DimRed_EOM}, if solved, determine the set of functions
\begin{align*}
 \left\{U,V_1,\ldots,V_{d-3},\gamma,\psi \right\}.
\end{align*}
Constructing the metric \Eqref{DimRed_metric} and scalar field \Eqref{DimRed_scalar} using these functions then determines a solution to the Einstein-scalar equations. If we have a known set of solutions, the equations \Eqref{DimRed_EOM} affords the following solution-generating procedures.
\begin{enumerate}
 \item \textbf{Scalar multipolar extensions.} The procedure to generate scalar multipole extensions \cite{Eris:1976xj,Herdeiro:2024oxn} can be applied in the present higher dimensional case. In particular, given a vacuum solution $\{U,V_1,\ldots,V_{d-3},\gamma,\psi=0\}$. One can include a scalar field $\psi'$ by taking a solution of $\vec{\nabla}^2\psi'=0$ and construct a function $\mu(\rho,z)$ by integrating
 \begin{subequations}\label{mu_integrate}
 \begin{align}
  \partial_\rho\mu&=\frac{d-2}{d-3}\sbrac{\brac{\partial_\rho\psi'}^2-\brac{\partial_z\psi'}^2},\\
  \partial_z\mu&=\frac{d-2}{d-3}\rho \partial_\rho\psi'\partial_z\psi'.
 \end{align}
 \end{subequations}
 Then, the function $\gamma'=\gamma+\mu$, along with the previous $U$ and $V_k$'s will be a solution to \Eqref{DimRed_EOM}. In other words, the new solution is
 \begin{align}
  \left\{U,V_1,\ldots,V_{d-3},\gamma+\mu,\psi'\right\}.
 \end{align}

 A general axisymmetric solution to the Laplace equation can be written in a multipolar expansion
 \begin{align}
  \psi'(\rho,z)=\sum_{l=0}^\infty\brac{\frac{a_l}{R^{l+1}}+b_lR^l}P_l\brac{\textstyle{\frac{z}{R}}},\label{multipolar_psi_soln}
 \end{align}
 for constant coefficients $a_l$ and $b_l$. Here $P_l(x)$ is the $l$-th Legendre polynomial and $R=\sqrt{\rho^2+z^2}$. The function $\mu$ is the integrated in Eq.~\Eqref{mu_integrate} to give
 \begin{align}
  \mu&=\sum_{l,m=0}^\infty\sbrac{\frac{(l+1)(m+1)a_la_m}{(l+m+2)R^{l+m+2}}\brac{P_{l+1}P_{m+1} - P_lP_m}}\nonumber\\
   &\hspace{1cm}+\sum_{l,m=1}^\infty\sbrac{\frac{lmb_lb_mR^{l+m}}{l+m}\brac{P_lP_m-P_{l-1}P_{m-1}}}. \label{multipolar_mu_soln}
 \end{align}

 \item \textbf{$SO(2)$ symmetry.} The equations in \Eqref{DimRed_EOM} are invariant under Buchdahl's $SO(2)$ transformation \cite{Buchdahl:1956zz,Buchdahl:1959nk,Abdolrahimi:2009dc,Barrientos:2024uuq}
 \begin{align}
   U\mapsto U\cos\beta-\psi\sin\beta,\quad \psi\mapsto U\sin\beta+\psi\cos\beta,
 \end{align}
 for some real parameter $\beta$. Therefore, given a solution with $U=U_0$ and $\psi=\psi_0$  to Eq.~\Eqref{DimRed_EOM}, one can generate a new solution with
 \begin{align}
  U=U_0\cos\beta-\psi_0\sin\beta\quad\mbox{ and }\quad \psi=U_0\sin\beta+\psi_0\cos\beta. \label{O2_transform}
 \end{align}
 When applied to the Schwarzschild solution, one obtains the Fisher--Janis--Newman--Winicour (FJNW) naked singularity \cite{Fisher:1948yn,Janis:1968zz}. Note that the $SO(2)$ transformation procedure only requires explicit use of one of the Killing vectors. (Namely the one we singled out by taking $U_{d-2}=U$.) Indeed, the procedure to derive the FJNW solution was previously applied in \cite{Abdolrahimi:2009dc,Lim:2017dqw} without requiring the existence of the other $d-3$ Killing vectors. Also, note that for this transformation to be available, it was important to cast the metric in the form \Eqref{DimRed_metric}, which releases $U$ from the determinant constraint \Eqref{DimRed_constraint}, making it free to participate in the $SO(2)$ transformation.
\end{enumerate}

\section{Applications of the scalar-generating procedures} \label{sec_Applications}

In this section we will apply the solution-generating procedure to vacuum seeds that can be brought into the generalized Weyl form. For the multipolar extensions we consider specific $l$-modes of \Eqref{multipolar_psi_soln}. Within each mode, there is the decaying and growing radial parts with coefficient $a_l$ and $b_l$, respectively. \RefOne{As was noted in \cite{Herdeiro:2024oxn}, the decaying solutions tend to turn horizons into singularities. These decaying solutions are contained in a theorem by Chase \cite{Chase:1970omy} which states that massless scalar fields in static, asymptotically flat spacetimes become singular at an event horizon. (See also \cite{Bekenstein:1996pn}.)} On the other hand the black hole horizons with growing multipoles remain horizons, but at the cost of introducing curvature singularities at infinity. To the scalar multipolar solutions, we apply Buchdahl's $SO(2)$ transformation to further extend the solution. The final outcomes of the procedure are then verified by directly checking the Einstein--scalar equations \Eqref{EinsteinScalar_EOM}.

\subsection{The four-dimensional Schwarzschild solution}

In its familiar form, the Schwarzschild solution \cite{Schwarzschild:1916ae} in four dimensions is
\begin{subequations}
\begin{align}
 \dif s^2&=-f(r)\dif t^2+f(r)^{-1}\dif r^2+r^2\brac{\dif\theta^2+\sin^2\theta\,\dif\phi^2},\\
     f(r)&=1-\frac{2m}{r},
\end{align}
\end{subequations}
where $m$ parametrizes the mass of the black hole. To bring it into the form of \Eqref{DimRed_metric}, we rearrange the terms in the metric to find
\begin{align*}
 \dif s^2&=-f\dif t^2+f^{-1}\cdot f\brac{r^2\sin^2\theta\,\dif\phi^2+\frac{\dif r^2}{f}+r^2\dif\theta^2}.
\end{align*}
With $\epsilon=-1$ and $d=4$, we then identify
\begin{align}
 \expo{U}=f^{1/2},\quad \expo{V_1}=f^{1/2}r\sin\theta.
\end{align}
The transformation into $(\rho,z)$-coordinates are given by \cite{Griffiths:2009dfa}
\begin{align}
 \rho=\sqrt{f}r\sin\theta,\quad z=(r-m)\cos\theta.
\end{align}

Let us first consider adding a scalar decaying monopole $l=0$. To Eq.~\Eqref{multipolar_psi_soln} we set $a_l=b_l=0$ for all $l\neq0$ along with $b_0=0$. Writing $a_0=a$ one finds \cite{Herdeiro:2024oxn}
\begin{align}
 \mu=-\frac{a^2\rho^2}{2(\rho^2+z^2)^2},\quad \psi=\frac{a}{\sqrt{\rho^2+z^2}}.
\end{align}
Constructing the metric with these functions we obtain
\begin{subequations}\label{soln_4dSch_decaying_monopole}
\begin{align}
 \dif s^2&=-f\dif t^2+\expo{-\frac{a^2fr^2\sin^2\theta}{[fr^2\sin^2\theta(r-m)^2\cos^2\theta]^2}}\brac{\frac{\dif r^2}{f}+r^2\dif\theta^2}+r^2\sin^2\theta\,\dif\phi^2,\\
 \varphi&=\frac{\sqrt{2}a }{\sqrt{(r^2-2mr)\sin^2\theta+(r-m)^2\cos^2\theta}}.
\end{align}
\end{subequations}
This solution was given in Eq.~(23) in Herdeiro's paper \cite{Herdeiro:2024oxn}. From this we can generate a new solution by applying the $SO(2)$ transformation \Eqref{O2_transform}, which results in
\begin{subequations} \label{soln_4dFJNW_decaying_monopole}
\begin{align}
 \dif s^2&=-f^\alpha\expo{-2\sqrt{1-\mu^2}\psi_0}\dif t^2+f^{1-\alpha}\expo{2\sqrt{1-\alpha^2}\psi_0}\sbrac{\expo{2\mu}\brac{\frac{\dif r^2}{f}+r^2\dif\theta^2}+r^2\sin^2\theta\,\dif\phi^2},\\
 \varphi&=\sqrt{2}\brac{\alpha\psi_0+\frac{\sqrt{1-\alpha^2}}{2}\ln f},\quad f=1-\frac{2m}{r},\\
 \quad\psi_0&=\frac{a}{\sqrt{(r^2-2mr)\sin^2\theta+(r-m)^2\cos^2\theta}},\quad\mu=\frac{a^2fr^2\sin^2\theta}{2[fr^2\sin^2\theta+(r-m)^2\cos^2\theta]^2},
\end{align}
\end{subequations}
where we have used $\alpha=\cos\beta$ for the quantity parametrizing the $SO(2)$ action. This is a solution described by three parameters $m$, $a$, and $\alpha$. Setting $\alpha=1$ recovers \Eqref{soln_4dSch_decaying_monopole}, while setting $a=0$ reduces to the FJNW solution. There is a curvature singularity at $r=2m$, which already occurs in the FJNW case. This solution is asymptotically flat as the metric approaches Minkowski in the limit $r\rightarrow\infty$.

Next, we consider a scalar growing dipole with $l=1$ and $a_1=0$. Letting $b_1=b$, one finds
\begin{align}
 \mu=-\half b^2\rho^2,\quad \psi=bz, \label{growing_dipole}
\end{align}
Reconstructing the metric and scalar field, we obtain
\begin{subequations}\label{soln_4dSch_growing_dipole}
\begin{align}
 \dif s^2&=-f\dif t^2+\expo{-b^2fr^2\sin^2\theta}\brac{\frac{\dif r^2}{f}+r^2\dif\theta^2}+r^2\sin^2\theta\,\dif\phi^2,\\
 \varphi&=\sqrt{2}b(r-m)\cos\theta.
\end{align}
\end{subequations}
This was given in Eq.~(36) of Herdeiro \cite{Herdeiro:2024oxn}. In the limit $m\rightarrow0$, we obtain the scalar counterpart to the Melvin universe \cite{Cardoso:2024yrb}. For general $m$ and $b$, we shall refer to this solution as the \emph{scalar counterpart to the Schwarzschild--Melvin solution}. From this we continue to generate a new solution by applying the $SO(2)$ transformation \Eqref{O2_transform}. The result is
\begin{subequations}\label{soln_4dFJNW_growing_dipole}
\begin{align}
 \dif s^2&=-f^{\alpha}\expo{-2\sqrt{1-\alpha^2}b(r-m)\cos\theta}\dif t^2+f^{1-\alpha}\expo{2\sqrt{1-\alpha^2}b(r-m)\cos\theta}\nonumber\\
 &\hspace{1cm}\times\sbrac{\expo{-b^2fr^2\sin^2\theta}\brac{\frac{\dif r^2}{f}+r^2\dif\theta^2}+r^2\sin^2\theta\,\dif\phi^2},\\
 \varphi&=\sqrt{2}\sbrac{\alpha b(r-m)\cos\theta+\half\sqrt{1-\alpha^2}\ln f},\quad f=1-\frac{2m}{r},
\end{align}
\end{subequations}
where, as before, we have used $\alpha=\cos\beta$ as the $SO(2)$ parameter. It is straightforward to see that setting $b=0$ recovers the FJNW solution and $\alpha=1$ recovers Herdeiro's growing dipole solution \Eqref{soln_4dSch_growing_dipole}.

Next we consider the decaying dipole, $l=1$ $a_1=a$ and all other $a_l$ and $b_l$ are zero. We obtain
\begin{align}
 \mu=\frac{a^2\rho^2(\rho^2-8z^2)}{4(\rho^2+z^2)^4},\quad \psi=\frac{az}{(\rho^2+z^2)^{3/2}}.
\end{align}
This results in the solution
\begin{subequations}\label{soln_4dSch_decaying_dipole}
\begin{align}
 \dif s^2&=-f\dif t^2+\expo{2\mu}\brac{\frac{\dif r^2}{f}+r^2\dif\theta^2}+r^2\sin^2\theta\,\dif\phi^2,\\
 \varphi&=\frac{\sqrt{2}a(r-m)\cos\theta}{[fr^2\sin^2\theta+(r-m)^2\cos^2\theta]^{3/2}},\quad\mu=\frac{a^2fr^2\sin^2\theta\sbrac{fr^2\sin^2\theta-8(r-m)^2\cos^2\theta}}{4\sbrac{fr^2\sin^2\theta+(r-m)^2\cos^2\theta}^4},
\end{align}
\end{subequations}
which is Eq.~(32) of Herdeiro \cite{Herdeiro:2024oxn}. Applying the $O(2)$ transformation to this solution, we get
\begin{subequations}
\begin{align}
 \dif s^2&=-f^\alpha\expo{-2\sqrt{1-\alpha^2}\psi_0}\dif t^2+f^{1-\alpha}\expo{2\sqrt{1-\alpha^2}\psi_0}\sbrac{\expo{2\mu}\brac{\frac{\dif r^2}{f}+r^2\dif\theta^2}+r^2\sin^2\theta\dif\phi^2},\\
 \varphi&=\sqrt{2}\psi_0,\quad f=1-\frac{2m}{r},\\
 \psi_0&=\frac{a(r-m)\cos\theta}{[fr^2\sin^2\theta+(r-m)^2\cos^2\theta]^{3/2}},
\end{align}
\end{subequations}
where $\mu$ is still as in Eq.~\Eqref{soln_4dSch_decaying_dipole}.

\subsection{The C-metric}

The C-metric \cite{Levi-Civita:1918,Weyl1919} is one of the earliest known exact solutions to Einstein equations, though it was much later that its interpretation as a spacetime describing a uniformly accelerating black hole was given \cite{Kinnersley:1970zw,Bonnor:1983,Griffiths:2009dfa,Griffiths:2006tk}. Here we shall use the factorized form \cite{Hong:2003gx}
\begin{subequations}
\begin{align}
 \dif s^2&=\frac{1}{A^2(x-y)^2}\sbrac{-Q(y)\dif t^2+\frac{\dif y^2}{Q(y)}+\frac{\dif x^2}{P(x)}+P(x)\dif\phi^2},\\
 Q(y)&=(y^2-1)(1+mAy),\quad P(x)=(1-x^2)(1+2mAx),
\end{align}
\end{subequations}
where $m$ and $A$ are the mass and acceleration parameters, respectively. In these coordinates, $y=-1$ and $y=-1/mA$ are the acceleration and black hole horizons, respectively, and $x=\pm 1$ are the polar axes of the spacetime.

Rearranging to bring it into the form \Eqref{DimRed_metric},
\begin{align*}
 \dif s^2&=-\frac{Q\dif t^2}{A^2(x-y)^2}+\frac{A^2(x-y)^2}{Q}\frac{Q}{A^4(x-y)^4}\sbrac{P\dif\phi^2+\brac{\frac{\dif y^2}{Q}+\frac{\dif x^2}{P}}},
\end{align*}
we identify
\begin{align}
 \expo{U}=\frac{\sqrt{Q}}{A(x-y)},\quad\expo{V_1}=\frac{QP}{A^2(x-y)^2}.
\end{align}
The transformation into $(\rho,z)$ coordinates is \cite{Bonnor:1983,Hong:2003gx,Emparan:2001wk}\footnote{See Appendix H of \cite{Harmark:2004rm} for detailed steps to obtain the Weyl form of metrics with C-metric-like coordinates.}
\begin{align}
 \rho=\expo{V_1}=\frac{\sqrt{QP}}{A^2(x-y)^2},\quad z=\frac{(1-xy)(1+\half mA(x+y))}{A^2(x-y)^2}.
\end{align}
Suppose we add a scalar growing dipole to this solution. Then $\mu$ and $\psi$ are the same as in the Schwarzschild case \Eqref{growing_dipole}, except that the interpretations of the $\rho$ and $z$ coordinates are different. The C-metric dressed with a scalar growing dipole is
\begin{subequations}
\begin{align}
 \dif s^2&=\frac{1}{A^2(x-y)^2}\sbrac{-Q\dif t^2+\expo{2\mu}\brac{\frac{\dif y^2}{Q}+\frac{\dif x^2}{P}}+P\dif x^2},\\
 Q&=(y^2-1)(1+mAy),\quad P=(1-x^2)(1+2mAx),\\
 \varphi&=\sqrt{2}b\frac{(1-xy)(1+mA(x+y)/2)}{A^2(x-y)^2},\quad\mu=-\frac{b^2PQ}{2A^4(x-y)^4}.
\end{align}
\end{subequations}
Applying the Buchdahl transformation \Eqref{O2_transform}, we get
\begin{subequations}
\begin{align}
 \dif s^2&=-\frac{Q^\alpha}{A^{2\alpha}(x-y)^{2\alpha}}\expo{-2\sqrt{1-\alpha^2}\psi_0}\dif t^2\nonumber\\
  &\quad+\frac{Q^{1-\alpha}}{A^{4-2\alpha}(x-y)^{4-2\alpha}}\expo{2\sqrt{1-\alpha^2}\psi_0}\sbrac{P\dif\phi^2+\expo{2\mu}\brac{\frac{\dif y^2}{Q}+\frac{\dif x^2}{P}}},\\
  Q&=(y^2-1)(1+mAy),\quad P=(1-x^2)(1+2mAx),\\
 \varphi&=\sqrt{2}\psi_0,\quad\mu=-\frac{b^2PQ}{2A^4(x-y)^4},\quad \psi_0=b\frac{(1-xy)(1+mA(x+y)/2)}{A^2(x-y)^2}.
\end{align}
\end{subequations}
Setting $b=0$ recovers the accelerating FJNW metric recently studied in \cite{Anjomshoa:2025tdl}.

\subsection{Five-dimensional Schwarzschild--Tangherlini solution}

The $d$-dimensional generalization of the Schwarzschild solution was given by Tangherlini \cite{Tangherlini:1963bw}. However, aside from the $d=4$ solution considered above, only the $d=5$ case can be brought into the Weyl form with three commuting Killing vectors \cite{Emparan:2001wk}. The metric is
\begin{subequations}
\begin{align}
 \dif s^2&=-f(r)\dif t^2+f(r)^{-1}\dif r^2+r^2\brac{\dif\theta^2+\sin^2\theta\,\dif\phi^2+\cos^2\theta\,\dif\zeta^2},\\
 f(r)&=1-\frac{\mu}{r^2},
\end{align}
\end{subequations}
where $\mu$ is the mass parameter and $(\theta,\phi,\zeta)$ are the angular coordinates of a three-sphere. The three commuting Killing vectors are $\partial_t$, $\partial_\phi$, and $\partial_\zeta$. Rearranging the metric into the form \Eqref{DimRed_metric} with $d=5$,
\begin{align*}
 \dif s^2&=-f\dif t^2+f^{-1/2} f^{1/2}\brac{r^2\sin^2\theta\,\dif\phi^2+r^2\cos^2\theta\,\dif\zeta^2+f^{-1}\dif r^2+r^2\dif\theta^2},
\end{align*}
we identify
\begin{align}
 \expo{U}=f^{1/2},\quad\expo{V_1}=f^{1/4}r\sin\theta,\quad \expo{V_2}=f^{1/4}r\cos\theta.
\end{align}
From this we can read off
\begin{align}
 \rho=\half f^{1/2}r^2\sin2\theta,
\end{align}
and following Emparan and Reall \cite{Emparan:2001wk}, the transformation to obtain the $z$-coordinate is
\begin{align}
 z=\half\brac{1-\frac{\mu}{2r^2}}r^2\cos 2\theta.
\end{align}
Taking the growing dipole solution $l=1$ and $a_1=0$, we find
\begin{align}
 \mu=-\frac{3}{8}b^2\rho^2,\quad \psi=bz. \label{5d_growing_dipole}
\end{align}
This results in the solution
\begin{subequations}\label{soln_5dSch_growing_dipole}
\begin{align}
 \dif s^2&=-f\dif t^2+\expo{-\frac{3}{16}b^2fr^4\sin^22\theta}\brac{\frac{\dif r^2}{f}+r^2\dif\theta^2}+r^2\brac{\sin^2\theta\,\dif\phi^2+\cos^2\theta\,\dif\zeta^2}, \\
 \varphi&=\frac{\sqrt{6}}{4} b\brac{1-\frac{\mu}{2r^2}}r^2\cos2\theta.
\end{align}
\end{subequations}
This is the five-dimensional version of the scalar counterpart to Schwarzschild--Melvin. Further applying the $O(2)$ transformation \Eqref{O2_transform}, we get
\begin{subequations}\label{soln_5dFJNW_growing_dipole}
\begin{align}
 \dif s^2&=-f^\alpha\expo{-2\sqrt{1-\alpha^2}\psi_0}\dif t^2\nonumber\\
      &\hspace{0.5cm}+f^{(1-\alpha)/2}\expo{\sqrt{1-\alpha^2}\psi_0}\sbrac{\expo{-\frac{3}{16}b^2fr^4\sin^22\theta}\brac{\frac{\dif r^2}{f}+r^2\dif\theta^2}+r^2\brac{\sin^2\theta\,\dif\phi^2+\cos^2\theta\,\dif\zeta^2}},\\
 \varphi&=\frac{\sqrt{6}}{2}\brac{\alpha\psi_0+\half\sqrt{1-\alpha^2}\ln f},\quad f=1-\frac{\mu}{r^2},\quad \psi_0=\frac{b}{2}\brac{1-\frac{\mu}{2r^2}}r^2\cos2\theta.
\end{align}
\end{subequations}
We see that setting $\alpha=1$ recovers \Eqref{soln_5dSch_growing_dipole}, and $b=0$ reduces to the $d=5$ FJNW solution \cite{Abdolrahimi:2009dc}.

\subsection{Static black ring}

The black ring is a five-dimensional black hole solution with an event horizon of non-spherical topology of $S^1\times S^2$. Here we consider the static black ring was presented by Emparan and Reall \cite{Emparan:2001wk}, where the metric is
\begin{align}
 \dif s^2&=-\frac{F(x)}{F(y)}\dif t^2\nonumber\\
  &\quad+\frac{1}{A^2(x-y)^2}\sbrac{F(x)\brac{(y^2-1)\dif\psi^2+\frac{F(y)\dif y^2}{y^2-1}}+F(y)^2\brac{\frac{\dif x^2}{1-x^2}+\frac{(1-x^2)\dif\phi^2}{F(x)}}},
\end{align}
where $F(\xi)=1-m\xi$. The solution are parametrized by constants $m$ and $A$ with $0\leq m\leq 1$ and $A>0$. The $(x,y)$ coordinates take the domain $-1\leq x\leq 1$ and $y\leq-1$. To bring it into the form \Eqref{DimRed_metric} with $d=5$, we rearrange the metric into
\begin{align*}
 \dif s^2&=-\frac{F(x)}{F(y)}\dif t^2+\brac{\frac{F(x)}{F(y)}}^{-1/2}\brac{\frac{F(x)}{F(y)}}^{1/2}\bigg[\frac{F(x)(y^2-1)\dif\psi^2}{A^2(x-y)^2}+\frac{F(y)^2(1-x^2)}{A^2(x-y)^2F(x)}\dif\phi^2\\
  &\hspace{5cm}+\frac{1}{A^2(x-y)^2}\brac{\frac{F(x)F(y)\dif y^2}{y^2-1}+\frac{F(y)^2\dif x^2}{1-x^2}}\bigg],
\end{align*}
we identify
\begin{align}
 \expo{U}=\brac{\frac{F(x)}{F(y)}}^{1/2},\quad \expo{V_1}=\brac{\frac{F(x)}{F(y)}}^{1/2}\frac{\sqrt{F(x)(y^2-1)}}{A(x-y)},\quad\expo{V_2}=\brac{\frac{F(x)}{F(y)}}^{1/2}\frac{F(y)\sqrt{1-x^2}}{A(x-y)\sqrt{F(x)}}.
\end{align}
The transformation into the $\rho$ and $z$ coordinates are \cite{Emparan:2001wk}
\begin{align}
 \rho=\frac{\sqrt{F(x)F(y)(1-x^2)(y^2-1)}}{A^2(x-y)^2},\quad z=\frac{(1-xy)(F(x)+F(y))}{2A^2(x-y)^2}.\label{BR_canonical_coords}
\end{align}

In $(\rho,z)$ coordinates, the equations for the scalar multipoles are identical to \Eqref{5d_growing_dipole}, even if the interpretations of $(\rho,z)$ are different between the Schwarzschild--Tangherlini and black ring solutions. In particular, the black ring dressed with a scalar growing dipole is
\begin{subequations}
\begin{align}
 \dif s^2&=-\frac{F(x)}{F(y)}\dif t^2+\frac{1}{A^2(x-y)^2}\bigg[F(x)(y^2-1)\dif\psi^2\frac{F(y)^2(1-x^2)\dif\phi^2}{F(x)}\nonumber\\
  &\hspace{4cm}+\expo{2\mu}\brac{\frac{F(x)F(y)\dif y^2}{y^2-1}+\frac{F(y)^2\dif x^2}{1-x^2}}\bigg],\\
  \varphi&=\sqrt{2}bz,\quad \mu=-\frac{3}{8}b^2\rho^2, \label{BR_growing_dipole}
\end{align}
\end{subequations}
where $(\rho,z)$ are expressed in the black ring coordinates $(x,y)$ according to \Eqref{BR_canonical_coords}.

Performing the $SO(2)$ transformation, we obtain
\begin{subequations}
\begin{align}
 \dif s^2&=-\brac{\frac{F(x)}{F(y)}}^\alpha\expo{-2\sqrt{1-\alpha^2}\psi_0}\dif t^2+\brac{\frac{F(x)}{F(y)}}^{(1-\alpha)/2}\expo{\sqrt{1-\alpha^2}\psi_0}\frac{1}{A^2(x-y)^2}\nonumber\\
   &\hspace{1cm}\times\bigg[F(x)(y^2-1)\dif\psi^2+\frac{F(y)^2(1-x^2)\dif\phi^2}{F(x)}+\expo{2\mu}\brac{\frac{F(x)F(y)\dif y^2}{y^2-1}+\frac{F(y)^2\dif x^2}{1-x^2}}\bigg],\\
   \varphi&=\sqrt{\frac{3}{2}}\brac{\alpha\psi_0+\frac{\sqrt{1-\alpha^2}}{2}\ln\frac{F(x)}{F(y)}},\\
   \psi_0&=\frac{b(1-xy)(F(x)+F(y))}{2A^2(x-y)^2},\quad F(\xi)=1-m\xi,
\end{align}
\end{subequations}
where $\mu$ is still as given in Eq.~\Eqref{BR_growing_dipole}. Setting $b=0$, we have the  ring version of the FJNW solution.

\section{Analysis of selected scalar multipolar solutions} \label{sec_Analysis}

In this section we select two solutions of potential physical interest to study some of its properties. The first is the scalar counterpart Schwarzschild--Melvin. The solution and the presence of curvature singularities have already been pointed out in \cite{Cardoso:2024yrb,Herdeiro:2024oxn}. Here we consider its causal structure and quasilocal energy. The second solution we study is the FJNW solution dressed with a decaying scalar monopole. Similar to the FJNW solution, there is a curvature singularity at its ``Schwarzschild radius'' $r=2m$, is asymptotically flat, and is sourced by a massless scalar field. Therefore we could perhaps regard this solution on equal footing with the original FJNW solution in terms of physical and observational interest.

\subsection{Scalar counterpart of the Schwarzschild--Melvin solution}

The scalar counterpart of the Schwarzschild--Melvin \cite{Cardoso:2024yrb,Herdeiro:2024oxn} was written above in \Eqref{soln_4dSch_growing_dipole}, and we rewrite it here for convenience:
\begin{subequations} \label{ScalarMelvin_soln}
\begin{align}
 \dif s^2&=-f\dif t^2+\expo{-b^2fr^2\sin^2\theta}\brac{\frac{\dif r^2}{f}+r^2\dif\theta^2}+r^2\sin^2\theta\dif\phi^2,\\
  \varphi&=\sqrt{2}b(r-m)\cos\theta,\quad f=1-\frac{2m}{r}.
\end{align}
\end{subequations}
Calculating some of its curvature invariants, we get
\begin{align}
 R&=\frac{2b^2\expo{b^2fr^2\sin^2\theta}}{r^2}\brac{r^2+m^2-2mr-m^2\cos^2\theta},\\
 R_{\mu\nu}R^{\mu\nu}&=\frac{4b^4\expo{2b^2fr^2\sin^2\theta}}{r^4}\big[r^4+m^4-4mr^3-4m^3r+6m^2r^2\nonumber\\
  &\hspace{1cm}+m^2\cos^2\theta\brac{m^2\cos^2\theta-2r^2+4mr-2m^2}\big],\\
  R_{\mu\nu\rho\sigma}R^{\mu\nu\rho\sigma}&=\frac{4\expo{2b^2fr^2\sin^2\theta}}{r^6}\big[12m^2+3r^6b^4+11r^2b^4m^4-32r^3b^4m^3-20r^2b^2m^2-16r^5b^4m\nonumber\\
   &\hspace{1cm}+16rb^2m^3+4mr^3b^2+34r^4b^4m^2+2b^2mr\cos^2\theta\big(18mr+18r^2b^2m^2+2r^4b^2\nonumber\\
   &\hspace{1cm}-8m^2-11rb^2m^3-11mr^3b^2-6r^2 \big)+b^4m^3r^2\cos^4\theta(11m-4r) \big].
\end{align}
We see that there are curvature singularities at $r\rightarrow0$ and $r\rightarrow\infty$. Clearly, we can check that the case $b=0$ reduces to the Schwarzschild results, $R=0$, $R_{\mu\nu}R^{\mu\nu}=0$, and $R_{\mu\nu\rho\sigma}R^{\mu\nu\rho\sigma}=\frac{48m^2}{r^6}$.

A Killing vector of this spacetime is $\xi^\mu=(C,0,0,0)$
for arbitrary constant $C$ which is timelike in the patch $2m<r<\infty$. The value of $C$ can be chosen to fix its normalization, with norm-squared $\xi_\mu\xi^\mu=-Cf(r)$. So we see that the surface $r=2m$ is a null surface corresponding to a Killing horizon. Since Killing horizons and event horizons coincide for static spacetimes, $r=2m$ remains an event horizon, just as in the Schwarzschild case. Leaving the normalization of the Killing vector yet unspecified, the surface gravity at $r=2m$ is
\begin{align}
 \kappa=\lim_{r\rightarrow2m}\sqrt{-\half\nabla_\mu\xi_\nu\nabla^\mu\xi^\nu}=\frac{C}{4m}, \label{ScalarMelvin_kappa}
\end{align}
which is independent of the scalar parameter $b$, and agrees with the Schwarzschild surface gravity up to the arbitrary normalization factor $C$.

Turning to the causal structure of the spacetime, we consider constant-$\theta$ and -$\phi$ sections of the metric
\begin{align}
 \dif s^2_2&=f\brac{-\dif t^2+\dif r_*^2}=-f\dif v\dif u, \label{ScalarMelvin_CausalMetric}
\end{align}
where in the first equality the metric is expressed in tortoise coordinates
\begin{align}
 r_*=\int\frac{\expo{-b^2fr^2\sin^2\theta/2}}{f}\dif r, \label{ScalarMelvin_tortoise}
\end{align}
and in the second it is expressed in double null coordinates
\begin{align}
 v=t+r_*,\quad u=t-r_*.
\end{align}
In the neighbourhood of the horizon, the function $f$ has the expansion
\begin{align}
 f\simeq f'(2m)(r-2m)+\ldots,
\end{align}
where $f'(2m)=\frac{1}{2m}$. Then in regions close to $r=2m$,
\begin{align}
 \frac{\expo{-b^2fr^2\sin^2\theta/2}}{f}&\simeq \brac{1+\frac{2m}{r-2m}}\sbrac{1-\frac{b^2r^2}{4m}\sin^2\theta(r-2m)+\ldots}\nonumber\\
   &\simeq1-\frac{b^2}{2}r^2\sin^2\theta+\frac{2m}{r-2m}+\ldots\nonumber
\end{align}
By integrating this expression, we see that the tortoise coordinate in the neighbourhood of the horizon is
\begin{align}
 r_*&\simeq r-\frac{b^2r^3}{6}\sin^2\theta+\frac{1}{2\kappa_0}\ln\left|\frac{r}{2m}-1 \right|+\ldots,
\end{align}
where $\kappa_0=\frac{1}{4m}$, is the surface gravity \Eqref{ScalarMelvin_kappa} if the choice $C=1$ is taken for the normalization of the timelike Killing vector. This shows that, similar to Schwarzschild, the tortoise coordinate $r_*$ has a logarithmic singularity at the horizon. We can then apply Kruskal's null coordinates
\begin{align}
 V=\expo{\kappa_0 v},\quad U=-\expo{-\kappa_0 u}
\end{align}
and the metric is transformed to
\begin{align}
 \dif s^2_2&\simeq-\frac{r}{2m\kappa_0^2}\expo{-2\kappa_0\brac{r-\frac{1}{6}b^2r^3\sin^2\theta}}\dif V\dif U
\end{align}
in the neighbourhood of the horizon, and is regular across $r=2m$.

\begin{figure}
 \centering
 \includegraphics{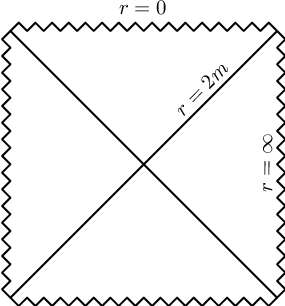}
 \caption{Conformal diagram of the scalar counterpart to the Schwarzschild--Melvin solution. The jagged lines indicate curvature singularities, and the diagonal lines represent the $r=2m$ event horizon.}
 \label{fig_ScalarMelvin_ConformalDiagram}
\end{figure}

The behavior of surfaces of constant $r$ can be seen from $UV=-\expo{2\kappa_0 r_*}$, which are hyperbolae in the $(U,V)$ plane. The locations of the singularities $r\rightarrow\infty$ and $r\rightarrow 0$ correspond to finite $r_*$, with an event horizon between them. Therefore it can be inferred that the curvature singularities are timelike and spacelike, respectively. To construct the conformal diagram we further introduce appropriate compactified coordinates and resulting conformal diagram is shown in Fig.~\ref{fig_ScalarMelvin_ConformalDiagram}.

\begin{figure}
 \centering
 \includegraphics{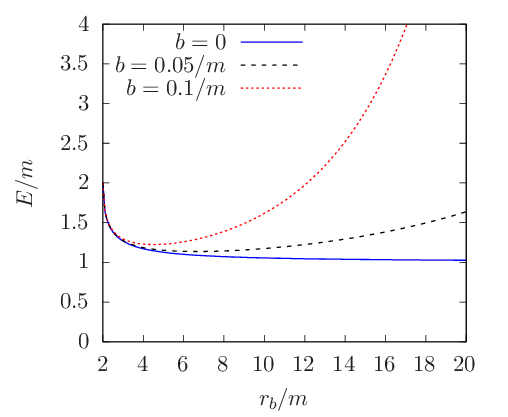}
 \caption{Quasilocal energy $E(r_b)$ against boundary radius $r_b$ for the scalar Schwarzschild--Melvin spacetime.}
 \label{fig_ScalarMelvin_Quasilocal}
\end{figure}

Because the spacetime is asymptotic to a curvature singularity at $r\rightarrow\infty$, the definitions of mass valid for asymptotically flat spacetimes cannot be applied. Nevertheless, we can attempt to determine the mass of the black hole inside a scalar-Melvin universe by applying the prescription of Brown and York \cite{Brown:1992br}. We choose a timelike boundary $\Sigma$ where $r=r_b$ and consider the energy content $E(r_b)$ contained within. Take $n^\mu=(0,\expo{b^2f(r_b)r_b^2\sin^2\theta/2}\sqrt{f(r_b)},0,0)$ to be the unit normal at the boundary $\Sigma$, and the induced metric on $\Sigma$ is
\begin{align}
 h_{ab}\dif y^a\dif y^b=-f(r_b)\dif t^2+\expo{-b^2f(r_b)r_b^2\sin^2\theta}r_b^2\dif\theta^2+r_b^2\sin^2\theta\dif\phi^2,
\end{align}
where $y^a$ are coordinates on $\Sigma$. The surface stress tensor is
\begin{align}
 T_{ab}=-\frac{1}{8\pi}\brac{K_{ab}-Kh_{ab}},
\end{align}
where $K_{ab}=e_a^\mu e_b^\nu\nabla_\mu n_\nu$ is the extrinsic curvature, and $e_a^\mu=\frac{\partial x^\mu}{\partial y^b}$ is the pullback/pushforward operator between $\Sigma$ and the bulk. We choose to measure the energy against a background of a horizonless $(m=0)$ scalar-Melvin universe, so that our metric is embedded in background with metric
\begin{align}
 \dif s_0^2&=-\dif\tau+\expo{-b'^2r^2\sin^2\theta}\brac{\dif r^2+r^2\dif\theta^2}+r^2\sin^2\theta\dif\phi^2
\end{align}
at $r>r_b$. To ensure the continuity of the metric across the boundary, the $\tau$ coordinate is scaled such that $\tau=\sqrt{f(r_b)}t$, and the scalar parameter is fixed to be $b'=b\sqrt{f(r_b)}$. The surface stress tensor on $\Sigma$ is \cite{Brown:1992br}
\begin{align}
 \mathcal{T}_{ab}=T_{ab}-{^0T_{ab}},
\end{align}
where ${^0T_{ab}}$ is the surface stress tensor of the background. Then the quasilocal energy is \cite{Brown:1992br} $E(r_b)=\oint_{\mathcal{B}}\dif^2\theta\sqrt{\gamma}\;\xi^a\xi^b\mathcal{T}_{ab}$ where $\xi_a$ is the timelike Killing vector pushed to $\Sigma$ and $\mathcal{B}$ is the spacelike surface normal to $\xi^a$ with induced metric $\gamma_{AB}$. The result is
\begin{align}
 E(r_b)=\half\int_0^\pi\dif\theta\,\sin\theta\;r_b H(r_b,\theta)^{-1/2}\sbrac{1+\frac{r_b^2H_0'(r_b,\theta)}{4H_0(r_b,\theta)}-f(r_b)^{1/2}\brac{1+\frac{r_b^2H'(r_b,\theta)}{4H(r_b,\theta)}}}, \label{ScalarMelvin_QuasilocalIntegration}
\end{align}
where we have denoted
\begin{align*}
 H(r,\theta)=\expo{-b^2f(r)r^2\sin^2\theta},\quad H_0(r,\theta)=H(r,\theta)=\expo{-b'^2r^2\sin^2\theta}
\end{align*}
and primes denote derivatives with respect to $r$. Setting $b=0$ with $m\neq0$, we get
\begin{align}
 E(r_b)=r_b\sbrac{1-f(r_b)^{1/2}},
\end{align}
which is the Schwarzschild quasilocal energy. (See, for instance, Eq.~(6.14) in \cite{Brown:1992br}). On the other hand $E(r_b)$ vanishes identically for $m=0$, since the spacetime and its background are identical. For generic values of $b$ and $m\neq0$, the integration in Eq.~\Eqref{ScalarMelvin_QuasilocalIntegration} may be performed numerically. The values of $E(r_b)$ against $r_b$ is shown in Fig.~\ref{fig_ScalarMelvin_Quasilocal} for $b=0$, $0.05/m$, and $0.1/m$. We see that the presence of the scalar field raises the energy contribution of the black hole. Because of the factor $H(r_b,\theta)^{-1/2}$ in the integrand, $E(r_b)$ diverges as $r_b\rightarrow\infty$. It is perhaps expected that this description breaks down as the curvature singularity at infinity is approached.

\subsection{FJNW solution with a decaying monopole}

The FJNW solution with a decaying monopole was obtained in \Eqref{soln_4dFJNW_decaying_monopole}, which we rewrite here for convenience:
\begin{subequations} \label{soln_4dFJNW_decaying_monopole2}
\begin{align}
 \dif s^2&=-f^\alpha\expo{-2\sqrt{1-\alpha^2}\psi_0}\dif t^2+f^{1-\alpha}\expo{2\sqrt{1-\alpha^2}\psi_0}\sbrac{\expo{2\mu}\brac{\frac{\dif r^2}{f}+r^2\dif\theta^2}+r^2\sin^2\theta\,\dif\phi^2},\\
 \varphi&=\sqrt{2}\brac{\alpha\psi_0+\frac{\sqrt{1-\alpha^2}}{2}\ln f},\quad f=1-\frac{2m}{r},\\
 \quad\psi_0&=\frac{a}{\sqrt{(r^2-2mr)\sin^2\theta+(r-m)^2\cos^2\theta}},\quad\mu=\frac{a^2fr^2\sin^2\theta}{2[fr^2\sin^2\theta+(r-m)^2\cos^2\theta]^2},
\end{align}
\end{subequations}
We see that $\varphi$ and $\mu$ tend to zero and $f$ tends to $1$ as $r\rightarrow\infty$. Therefore the solution asymptotically approaches Minkowski.

The curvature invariants are too cumbersome to be displayed here, but $R$, $R_{\mu\nu}R^{\mu\nu}$, and $R_{\mu\nu\rho\sigma}R^{\mu\nu\rho\sigma}$ can be checked using computer algebra software\footnote{The author used MAPLE.} that there is a curvature singularity at $r=2m$. As was already pointed out by Herdeiro \cite{Herdeiro:2024oxn}, dressing the solution with decaying multpole solutions tends to turn the $r=2m$ Schwarzschild horizon into a curvature singularity. On the other hand, the $r=2m$ surface is already a curvature singularity in the original FJNW solution. Hence the inclusion of the decaying scalar monopole does not introduce any new singularities to the solution.

\begin{figure}
 \centering
 \includegraphics{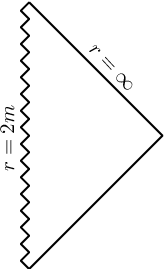}
 \caption{Conformal diagram for the FJNW spacetime with a scalar decaying monopole.}
 \label{fig_FisherMonopole_ConformalDiagram}
\end{figure}

To investigate its causal structure, we consider a constant-$\theta$ and -$\phi$ section of the metric
\begin{align}
 \dif s_2^2=-f^\alpha\expo{-2\sqrt{1-\alpha^2}\psi_0}\dif t^2+f^{-\alpha}\expo{2\mu+2\sqrt{1-\alpha^2}\psi_0}\dif r^2.
\end{align}
introducing tortoise coordinates
\begin{align}
 r_*=\int f^{-\alpha}\expo{\mu+2\sqrt{1-\alpha^2}\psi_0}\dif r,
\end{align}
the metric can now take the form
\begin{align}
 \dif s_2^2&=f^\alpha\expo{-2\sqrt{1-\alpha^2}\psi_0}\brac{-\dif t^2+\dif r_*^2}.
\end{align}
In the neighbourhood of the $r=2m$ singularity, the function $f$ has the expansion $f\simeq\frac{r-2m}{2m}+\ldots$ so the integral for the tortoise coordinate reads
\begin{align}
 r_*=\int\brac{\frac{r-2m}{4m}}^{-\alpha}\expo{\mu+2\sqrt{1-\alpha^2}\psi_0}\dif r+\ldots,
\end{align}
so that, with $0\leq\alpha<1$ the domain $2m<r<\infty$ corresponds to $r_*^0<r_*<\infty$ for some finite value of $r_*^0$. By introducing null coordinates $v=t+r_*$ and $u=t-r_*$ and compactifying, we conclude that the conformal diagram is as shown in Fig.~\ref{fig_FisherMonopole_ConformalDiagram}, which is the same as the original FJNW spacetime \cite{Martinez:2020hjm}.

As this spacetime is asymptotically Minkowski, we shall use the Komar prescription of the energy content of the spacetime,
\begin{align}
 E=-\frac{1}{8\pi}\int_{S_r} *\dif\xi,
\end{align}
where $*$ is the Hodge dual operation and $\dif\xi=\half(\partial_\mu\xi_\nu-\partial_\nu\xi_\mu)\dif x^\mu\wedge\dif x^\nu$ is the exterior derivative of the Killing one-form $\xi=\xi_\mu\dif x^\mu$, which we obtain by lowering the index of the timelike Killing vector $\xi^\mu=(1,0,0,0)$. (We choose to normalize the Killing vector to have unit norm at infinity.) The integral is performed over a two-sphere whose radius is taken to infinity, and we find
\begin{align}
 E=\alpha m,
\end{align}
which is independent of the scalar monopole parameter $a$, and is equal to the case of the FJNW spacetime. In other words, inclusion of the scalar monopole field does not change the Komar mass of the FJNW spacetime. This perhaps demonstrates a notion of non-uniqueness of  singular spacetimes with scalar fields, since we have a whole family of solutions parametrised by $b$, all of which are static, asymptotically flat and have the same (Komar) mass $\alpha m$.

\section{Inclusion of magnetic fields} \label{sec_EM}

We now introduce electromagnetic fields to our problem. The action now is
\begin{align}
 I=\frac{1}{16\pi}\int\dif^d x\sqrt{-g}\brac{R-F^2-\brac{\nabla\varphi}^2},
\end{align}
where $F^2=F_{\mu\nu}F^{\mu\nu}$ is the self contraction of the Faraday 2-form $F=\dif A$ with components $F_{\mu\nu}=\partial_\mu A_\nu-\partial_\nu A_\nu$ and $A=A_\mu\dif x^\mu$ is the gauge potential. By extremising the action, we get the Einstein--Maxwell--scalar equations
\begin{align}
 R_{\mu\nu}&=\nabla_\mu\varphi\nabla_\nu\varphi+2F_{\mu\lambda}{F^\lambda}_\nu-\frac{1}{d-2}F^2g_{\mu\nu},\quad \nabla_\mu F^{\mu\nu}=0, \quad\nabla^2\varphi=0. \label{EinsteinMaxwellScalar_EOM}
\end{align}
We take the same ansatz \Eqref{DimRed_metric} and \Eqref{DimRed_scalar} for the metric and scalar field. As for the electromagnetic field we write
\begin{align}
 A=\sqrt{\frac{d-2}{2(d-3)}}\chi\dif\sigma,
\end{align}
where $\chi$ is assumed to depend only on $\rho$ and $z$. Using the ansatz on Eq.~\Eqref{EinsteinMaxwellScalar_EOM}, the equations of motion are
\begin{subequations}
\begin{align}
 \vec{\nabla}^2U&+\epsilon\expo{-2U}\brac{\vec{\nabla}\chi}^2=0,\quad\vec{\nabla}^2V_k=0,\quad \vec{\nabla}\cdot\brac{\expo{-2U}\vec{\nabla}\chi}=0,\quad\vec{\nabla}^2\psi=0,\\
 \partial_\rho\gamma&=-\frac{1}{2\rho}+\frac{\rho}{2}\sum_{k=1}^{d-3}\sbrac{(\partial_\rho V_k)^2-(\partial_z V_k)^2}\nonumber\\
   &\hspace{1cm}+\frac{d-2}{d-3}\frac{\rho}{2}\sbrac{(\partial_\rho U)^2-(\partial_zU)^2+(\partial_\rho\psi)^2-(\partial_z\psi)^2+\epsilon\expo{-2U}\brac{(\partial_\rho\chi)^2-(\partial_z\chi)^2}},\\
   \partial_z\gamma&=\rho\sum_{k=1}^{d-3}\partial_\rho V_k\partial_z V_k+\frac{d-2}{d-3}\rho\brac{\partial_\rho U\partial_zU+\partial_\rho\psi\partial_z\psi+\epsilon\expo{-2U}\partial_\rho\chi\partial_z\chi},
\end{align}
\end{subequations}
where we are using the notation $\vec{\nabla}^2U=\sigma^{ab}\nabla_a\nabla_bU=\frac{1}{\sqrt{\sigma}}\partial_a\brac{\sqrt{\sigma}\sigma^{ab}\partial_bU}$, $\brac{\vec{\nabla}\chi}^2=\sigma^{ab}\partial_a\chi\partial_b\chi$, and $\vec{\nabla}\cdot\brac{\expo{-2U}\vec{\nabla}\chi}=\sigma^{ab}\nabla_a\brac{\expo{-2U}\nabla_b\chi}$ for inner products and derivative operations defined with respect to the unphysical auxiliary metric \Eqref{R3_aux}.

We see that the function $U$ couples to the potential $\chi$ and the system is no longer invariant under the Buchdahl's $SO(2)$ transformation \Eqref{O2_transform}. Nevertheless, this system is invariant under a Harrison-type transformation \cite{Harrison:1968,Dowker:1993bt,Ortaggio:2004kr}
\begin{align}
 U\rightarrow U'&=U-\ln\Lambda,\quad\chi\rightarrow\chi'=\Lambda^{-1}\sbrac{\chi+c\brac{\epsilon\expo{2U}+\chi^2}},\nonumber\\
 \Lambda&=\brac{1+c\chi^2}+\epsilon c^2\expo{2U},\label{Harrison_transform}
\end{align}
for a real constant $c$. It reduces to an identity operation when $c=0$. In other words, given a seed $(U,\chi)$ known to solve the equations of motion, a new set $(U',\chi')$ obtained as above will also be a solution. If $\partial_\sigma$ is a time-like Killing vector ($\epsilon=-1$, $\sigma=t$,) then $A=\sqrt{\frac{d-2}{2(d-3)}}\chi\,\dif t$ sources an electric field. On the other hand if $\partial_\sigma$ is an axial Killing vector ($\epsilon=+1$, $\sigma=\phi$), then $A$ sources a magnetic field. The latter case, in the absence of the scalar field, this is the transformation that generates the Melvin and Schwarzschild--Melvin and various other magnetized solutions \cite{Ernst:1976mzr,Dowker:1993bt,Galtsov:1998mhf,Yazadjiev:2005gs,Ortaggio:2004kr}. From the above equations we see that the terms involving $\varphi$ are decoupled from those with $U$ and $\chi$. So the Harrison-type transformation can be applied to solutions with $\varphi\neq0$. Or equivalently, the scalar-generating procedure  of adding scalar fields can be applied to a solution to the Einstein--Maxwell equations.

We now apply the Harrison-type transformation to the scalar counterpart of Schwarzschild--Melvin, whose solution was written above in \Eqref{ScalarMelvin_soln}. Choosing to perform the magnetising transformation, we bring the metric into the form \Eqref{DimRed_metric} with $\sigma=\phi$ to be the angular coordinate. We have $\epsilon=+1$ and the solution is written as
\begin{align}
 \dif s^2&=r^2\sin^2\theta\,\dif\phi^2+\brac{r^2\sin^2\theta}^{-1}\brac{r^2\sin^2\theta}\sbrac{-f\dif t^2+\expo{-b^2fr^2\sin^2\theta}\brac{\frac{\dif r^2}{f}+r^2\dif\theta^2}},\nonumber\\
 \varphi&=\sqrt{2}b(r-m)\cos\theta,
\end{align}
We then identify
\begin{align}
 \expo{U}=r\sin\theta,\quad \expo{V}=f,\quad\psi=b(r-m)\cos\theta,\quad\chi=0.
\end{align}
For the Harrison transformation, the seed is $(U,\chi)=(\ln r\sin\theta,0)$ is the same seed for Schwarzschild--Melvin without the scalar field. So the transformation \Eqref{Harrison_transform} goes through the same as the original Schwarzschild--Melvin and the result is
\begin{align}
 U'=\ln\frac{r\sin\theta}{\Lambda},\quad\chi'=\frac{cr^2\sin^2\theta}{\Lambda},
\end{align}
where $\Lambda=1+c^2r^2\sin^2\theta$. Writing $c=\half B$, then reconstructing the metric and the rest of the solution, it reads
\begin{subequations}
\begin{align}
 \dif s^2&=\Lambda^2\sbrac{-f\dif t^2+\expo{-b^2fr^2\sin^2\theta}\brac{\frac{\dif r^2}{f}+r^2\dif\theta^2}} + \frac{r^2\sin^2\theta}{\Lambda^2}\dif\phi^2,\\
 \chi&=\frac{Br^2\sin^2\theta}{2\Lambda}\,\dif\phi,\quad\varphi=\sqrt{2}b(r-m)\cos\theta,\\
 \Lambda&=1+\frac{1}{4}B^2r^2\sin^2\theta,\quad f=1-\frac{2m}{r}.
\end{align}
\end{subequations}
This solution can be verified by checking \Eqref{EinsteinMaxwellScalar_EOM} directly. We have now a single solution containing both the magnetic and scalar counterparts of the Schwarzschild--Melvin spacetime. The pure magnetic and pure scalar counterparts are recovered by taking $b=0$ and $B=0$, respectively.

We now explore some basic properties of the solution. The Ricci scalar is
\begin{align}
 R=\frac{32(r^2-2mr+m^2\sin^2\theta)b^2}{(4+B^2r^2\sin^2\theta)^2r^2}\expo{b^2fr^2\sin^2\theta}.
\end{align}
Other scalar invariants such as $R_{\mu\nu}R^{\mu\nu}$ and $R_{\mu\nu\rho\sigma}R^{\mu\nu\rho\sigma}$ are too cumbersome to be displayed here, but we can check that they are regular at $r=2m$, and there are curvature singularities at $r\rightarrow0$ and $r\rightarrow\infty$. In the patch $2m<r<\infty$, the spacetime has a timelike Killing vector $\xi^\mu=(C,0,0,0)$ for some constant $C$. Using the formula \Eqref{ScalarMelvin_kappa} for the present solution, the surface gravity at the horizon is the same, $\kappa=C/4m$.

The causal structure is also the same as in the $B=0$ case. To see this, consider the constant-$\theta$ and -$\phi$ section of the spacetime
\begin{align}
 \dif s_2^2&=\Lambda^2\sbrac{-f\dif t^2+\expo{-b^2fr^2\sin^2\theta}\frac{\dif r^2}{f}},
\end{align}
which is equivalent to \Eqref{ScalarMelvin_CausalMetric} up to a conformal transformation with $\Lambda$ being the conformal factor. Since null paths are invariant under conformal transformations, the light cone structure in the radial direction is the same as the $B=0$ case. Hence we conclude that the conformal diagram is also Fig.~\ref{fig_ScalarMelvin_ConformalDiagram}.

The magnetic flux is also unchanged from the $b=0$ case, as the form of the gauge potential $\chi$ is independent of $b$. Explicitly, the magnetic flux is $\Phi=\oint_C A_\mu\dif x^\mu$ where $C$ is a closed loop. Choosing the loop to have a radius $r_0$, we get
\begin{align}
 \Phi=\frac{\pi Br_0^2\sin^2\theta}{1+\frac{1}{4}B^2r_0^2\sin^2\theta},
\end{align}
which is the same flux as the usual Schwarzschild--Melvin case \cite{Ortaggio:2004kr} and is unaffected by the presence of the scalar field.

\section{Conclusion} \label{sec_Conclusion}

In this work we have considered black holes in scalar multipolar universes using the generalized Weyl metrics. The Schwarzschild(--Tangherlini) black hole in four and five dimensions, the C-metric, as well as the static black ring can be written in Weyl form and scalar multipolar extensions have been obtained. Some properties of the scalar counterpart to the Schwarzschild--Melvin solution are explored.

As was already pointed out in \cite{Herdeiro:2024oxn}, \RefOne{along with earlier results from \cite{Chase:1970omy,Bekenstein:1996pn}}, solutions with a decaying multipole tend to turn the black hole horizon into a curvature singularity whereas the singularity of the growing multipole occurs at infinity. In the latter case the black hole horizon is preserved. By applying the Buchdahl transformations to the scalar multipolar solution, we obtain a one-parameter extension to the multipolar solutions. Turning off the multipoles in these solutions recovers the FJNW metric. The Komar mass of the solution with a decaying monopole was calculated and is the same as the FJNW case. In other words, the scalar monopole part of the field does not contribute to the FJNW mass. This also demonstrates the non-uniqueness of singular spacetimes.

When electromagnetic fields are included with an appropriate ansatz for the gauge potential, one can find further solutions by making use of the Harrison transformations. Through this we were able to magnetize the scalar counterpart to the Schwarzschild--Melvin solution, thus producing a general solution which contains both the magnetic and scalar counterparts of Schwarzschild--Melvin. The Maxwell and scalar fields appear at different sectors of the solution. Therefore the many of its properties remain similar to its pure magnetic or pure scalar counterparts. For instance there is still a curvature singularity at infinity, the causal structure is the same as the pure scalar case, and the magnetic flux is the same as the pure magnetic case.

In the generalized Weyl construction, all $d-2$ Killing vectors are orthogonal, and hence the Lorentzian solutions considered here are static. Stationary or rotating solutions with scalar extensions have been considered in \cite{Barrientos:2024uuq,Stelea:2025ppj}. These results may potentially be extended further in higher dimensions using the canonical form introduced by Harmark \cite{Harmark:2004rm}, in which the $d-2$ Killing vectors are not necessarily orthogonal, and thus can capture stationary solutions. An example of a higher-dimensional stationary solution is the Myers--Perry black hole \cite{Myers:1986un}. \RefOne{The addition of scalar-multipolar fields to Myers--Perry was already done by Barrientos et al. in Refs.~\cite{Barrientos:2024uuq,Barrientos:2025abs}. Besides the Myers--Perry, in higher dimension one also has the rotating black rings \cite{Emparan:2001wn,Pomeransky:2006bd} for which these procedures can be applied, serving as a potential avenue for future work.}

\RefOne{Another potential avenue worth exploring may be the recent intriguing result which demonstrates a mapping a Skyrme--Einstein--Maxwell theory to an Einstein--Maxwell--scalar theory \cite{Canfora:2026wfi}. Due to this mapping, under certain conditions, exact solutions of the latter can be mapped to the former with a baryonic charge and magnetic field. Then, solution-generating procedures in the Einstein--Maxwell--scalar model can be used to find solutions in the Skyrme--Einstein--Maxwell counterparts. In Ref.~\cite{Barrientos:2026kdl}, this has used to obtain Melvin and Bertotti--Robinson solutions with baryonic charges. It then might the worth investigating whether other solutions (obtained here or elsewhere) can be applied to find other Skyrme--Einstein--Maxwell solutions.}

\section*{Acknowledgments}
Y.-K.~L is supported by Xiamen University Malaysia Research Fund (Grant no. XMUMRF/2021 -C8/IPHY/0001).

\bibliographystyle{canonical}

\bibliography{canonical}

\end{document}